\documentclass[useAMS,usenatbib,onecolumn]{mn2e}

\usepackage{graphicx}
\usepackage{amssymb}
\usepackage{natbib}
\usepackage{cite}

\bibpunct{(}{)}{;}{a}{}{,}
\usepackage{hyperref}

\begin{document}

\title[Shock-ionization in the EELR of 3C~305]{Shock-ionization in the Extended Emission-Line Region of 3C~305. The last piece of the (optical) puzzle.\thanks{Based on observations obtained at the Gemini Observatory, which is operated by the Association of Universities for
    Research in Astronomy, Inc., under a cooperative agreement with the
    NSF on behalf of the Gemini partnership: the National Science
    Foundation (United States), the Science and Technology Facilities
    Council (United Kingdom), the National Research Council (Canada),
    CONICYT (Chile), the Australian Research Council (Australia),
    Minist\'{e}rio da Ci\^{e}ncia, Tecnologia e Inova\c{c}\~{a}o (Brazil)
    and Ministerio de Ciencia, Tecnolog\'{i}a e Innovaci\'{o}n Productiva
    (Argentina)}
} 

\author[V. Reynaldi and C. Feinstein]{V. Reynaldi$^{1,2}$
 and C. Feinstein$^{1,2}$
\thanks{E-mail:vreynaldi@fcaglp.unlp.edu.ar (VR); cfeinstein@fcaglp.unlp.edu.ar (CF)}\\
$^{1}$Facultad de Ciencias Astron\'{o}micas y Geof\'{i}sicas, UNLP, Paseo del Bosque s/n, La Plata 1900, Argentina\\
$^{2}$Instituto de Astrof\'{i}sica de La Plata, CONICET, Argentina}


\date{}

\pagerange{\pageref{firstpage}--\pageref{lastpage}} \pubyear{}

\maketitle

\label{firstpage}

\begin{abstract}
We present new Gemini spectroscopical data of the Extended Emission-Line Region of 3C~305 radio galaxy in order to achieve the final answer of the long-standing question about the ionizing mechanism. The spectra show strong kinematic disturbances within the most intense line-emitting region. The relative intensities amongst the emission lines agree with the gas being shocked during the interaction of the powerful radio jets with the ambient medium. The emission from the recombination region acts as a very effective cooling mechanism, which is supported by the presence of a neutral outflow. However, the observed intensity is almost an order of magnitude lower than expected in a pure shock model. So auto-ionizing shock models, in low-density and low-abundance regime, are required in order to account for the observed emission within the region. This scenario also supports the hypothesis that the optical emitting gas and the X-ray plasma are in pressure balance.
\end{abstract}



\section{Introduction}

The ionizing mechanism on the optical emission-line regions of powerful radio galaxies is always an interesting topic to debate. The multiwavelength-emitter nature of these objects, the radiating power of the active nucleus, the role that jets might play, make them fascinating scenarios to test the scope of the two main likely mechanisms: nuclear photoionization and shock-ionization. Direct interaction between the radio jets and the optical-emitting gas has been reported in a wide sample of 3CR sources \citep{fei99,mar00,sol02,til05,chr06,har10}. However, sometimes the traces of such a violent process remain present within the gas even when those structures seem not related \citep{sol01,fei02,rey13}.

The interaction between the radio jets and the interstellar medium (ISM) in 3C~305 has been reported by several multiwavelength investigations, most of which have studied the strongly disturbed kinematics. Optical line-splitting has been used to show the violent processes that the gas within the Extended Emission-Line Region (EELR) is undergoing \citep{heck82,har12}. The earlier work of \citet{heck82} had already pointed out that the most complex and disordered kinematics of the EELR was located near the most intense radio features, and that the relation between these structures was not restricted to the kinematics. More recently, it was noticed that the optical gas is not the only ISM's phase in 3C~305 that the jet-cloud interaction can alter. H {\sc i} observations have been key to show that in such process: 1) the gas clouds are not destroyed by the pass of the relativistic jet, 2) the high-velocity neutral and ionized clouds are ejected by the same process, and 3) the ambient gas can cool in a very efficient way after being shocked \citep{mor05a}. The presence of the [Fe{\sc ii}]$\lambda$1.644$\mu$m infrared emission-line, which is spatially coincident with [O{\sc iii}]$\lambda5007$, also argues for a shocked medium \citep{jack03}. Towards the other side of the electromagnetic (EM) spectrum, it was shown that the X-ray emission is strongly related to the [O {\sc iii}] structures in this galaxy, not only by the morphological coincidence but also by their possible common origin: the X-ray emission is concentrated below $\sim$2~keV where most of the collisionally excited emission lines are formed \citep{bia06,mass09,har12}. 

All these self-consistent results reveal that radio (non-thermal), H{\sc i}-21cm, IR, optical and X-ray emission not only are overlaped but also are intrinsically related. The long-standing dichotomy between nuclear photoionization and jet-driven shock-ionization processes within the EELR in 3C~305 is ending now. 

In order to finally address this issue, we have obtained new optical spectra of the EELR in 3C~305 with unprecedented spatial and spectral resolution (Section~\ref{section:Data}). We have chosen to put the slit in roughly the jet direction (the difference is due to a rather higher surface brightness over the chosen direction; Section~\ref{section:Results}) because we want to explore the physical conditions of the EELR gas in the same place where the radio jets have been interacting with it. In Section~\ref{section:discussion} we discuss the main ionizing mechanism with the background of all previous results in every EM band. We also add the diagnostic analysis based upon the high quality emission-lines in our new long-slit spectra. Section~\ref{section:summary} collects all the pieces to finally solve the puzzle, which confirms that jet-induced shocks have triggered the EELR emission.

%
%

\section{ New Gemini/GMOS data.}
\label{section:Data}
We observed 3C~305 ($z=0.04164$) on April 2011 with the 8m-Gemini North Telescope as part of the GN-2011A-Q-66 program (PI:Feinstein). The GMOS facility was set up in long-slit mode, placing the 0.75 arcsec-width slit at PA=$45^\circ$ lying on the optical-UV emission region. We used the B600-G5307 grating, R$\sim$1700, which yields a resolution of 0.9~\AA~px$^{-1}$. One exposure of 2400~s was taken with 2$\times$2 binned CCD. Since the binned spectrum has a resolution of 0.1454~arcsec~px$^{-1}$, it gives a linear scale\footnote{$H_0$=73~km~s$^{-1}$~Mpc$^{-1}$,$\Omega_\mathrm{mat}$=0.27; $\Omega_{\Lambda}$=0.73} of about 124~pc~px$^{-1}$, without projection correction. Data were reduced following the usual steps of bias subtraction, flat-field correction, wavelength calibration and sky subtraction by using the {\sc gemini-gmos} package reduction tasks within {\sc iraf}. Cosmic ray (CR) rejection was not performed because the two available tasks ({\it gscrrej} from {\sc gemini} package, and the independent {\it lacos\_scpec} task \citep{dok01}) sometimes fail in the cleaning of the spectrum. We have found some cases where the CR hit is not removed; but worse, in other cases the tasks left a hole in the place of the CR rather than cleaning it. This failure is particulary dangerous when the CR hits the line profile. For this reason, we did not remove cosmic rays from the entire spectrum. However, they can be eliminated individually over the line profile once the Gaussian decomposition is being carried out, therefore they do not represent an obstacle to the measurements.

The spectral range covers from $\sim$3600\AA~  up to $\sim$5800\AA. The lines from oxygen ions dominate the spectrum. Table~\ref{table-1} lists the set of identified emission lines within both northeast and southwest EELR, the relative fluxes to that of H$\beta$ are also listed. The angular distances are measured with respect to the galactic centre, which was identified as the peak of intensity within the continuum emission in several featureless regions of the spectrum. The continuum emission drops below the noise in our GMOS spectrum at $\sim$1~arcsec from the galactic centre, so we have defined the central region as the inner 3 arcsec. In the following analysis the central region was excluded to prevent our measurement being affected by the stellar component. Many absortion features are also present within this region. 

\begin{figure}
\begin{center}
\includegraphics[scale=0.9,angle=0]{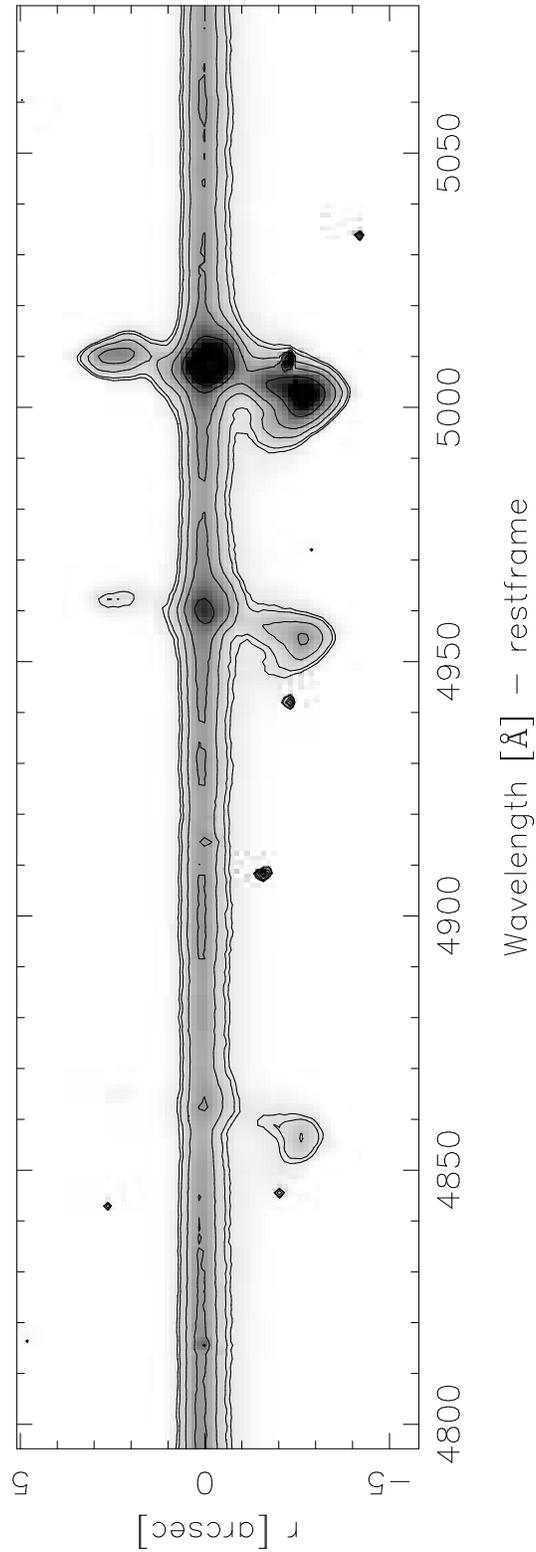}
\caption{A section of the long-slit Gemini spectrum. The lines are H$\beta$ and the doublet [O{\sc iii}]$\lambda \lambda 4959,5007$. The ordinate scale indicates the NE (negative) and SW (positive). The contour levels are shown only to enphasise the grey-scale contrast. There are tails of [O{\sc iii}]$\lambda5007$ low-brightness emission extending up to 15 arcsec NE and 9 arcsec SW that are not shown here. \label{figure-1}}
\end{center}
\end{figure}

Fig.~\ref{figure-1} shows the three most intense lines: H$\beta$ and the doublet [O{\sc iii}]$\lambda \lambda 4959,5007$ in grey scale, and reveals how complex the lines are. The contours have been plotted with the aim of emphasizing the line's intrinsic shapes. The dot-like feature superimposed on the [O{\sc iii}]$\lambda5007$ line at around 5010 \AA~ is not part of the line's structure but a cosmic ray hit. The [O{\sc iii}]$\lambda5007$ emission extends up to 15 arcsec towards the NE and 9 arcsec to the SW, altough it is not shown in this figure due to its low intensity. These tails are only detected in [O{\sc iii}]$\lambda5007$.

%
%

\section{Results}
\label{section:Results}
We can find almost all the emission concentrated inside $\pm$4~arcsec ($\sim$6.8~kpc) from the nucleus, but the tails of [O{\sc iii}]$\lambda5007$ emission reveal that the optical-emitting gas is extended as much as the X-ray emission \citep{har12}. The total extension of line emission along the slit direction (PA=$45^\circ$, north to east) is then 24 arcsec ($\sim$20~kpc). The radio axis location lies in roughly the same direction \citep[PA=$42^\circ$;][]{mor05a}. Peaks of line-emission are offset from the nuclear region, as was found by \citet{heck82}; there is a maximum of emission toward the NE at 2-3 arcsec, and a second weaker maximum at the same distance toward SW. The NE knot in our spectrum lies near the NE radio hot spot, there is also a bright X-ray feature at the same place \citep[see figs.3-4 from][]{har12}.

\begin{figure}
\begin{center}
\includegraphics[scale=0.8,angle=0]{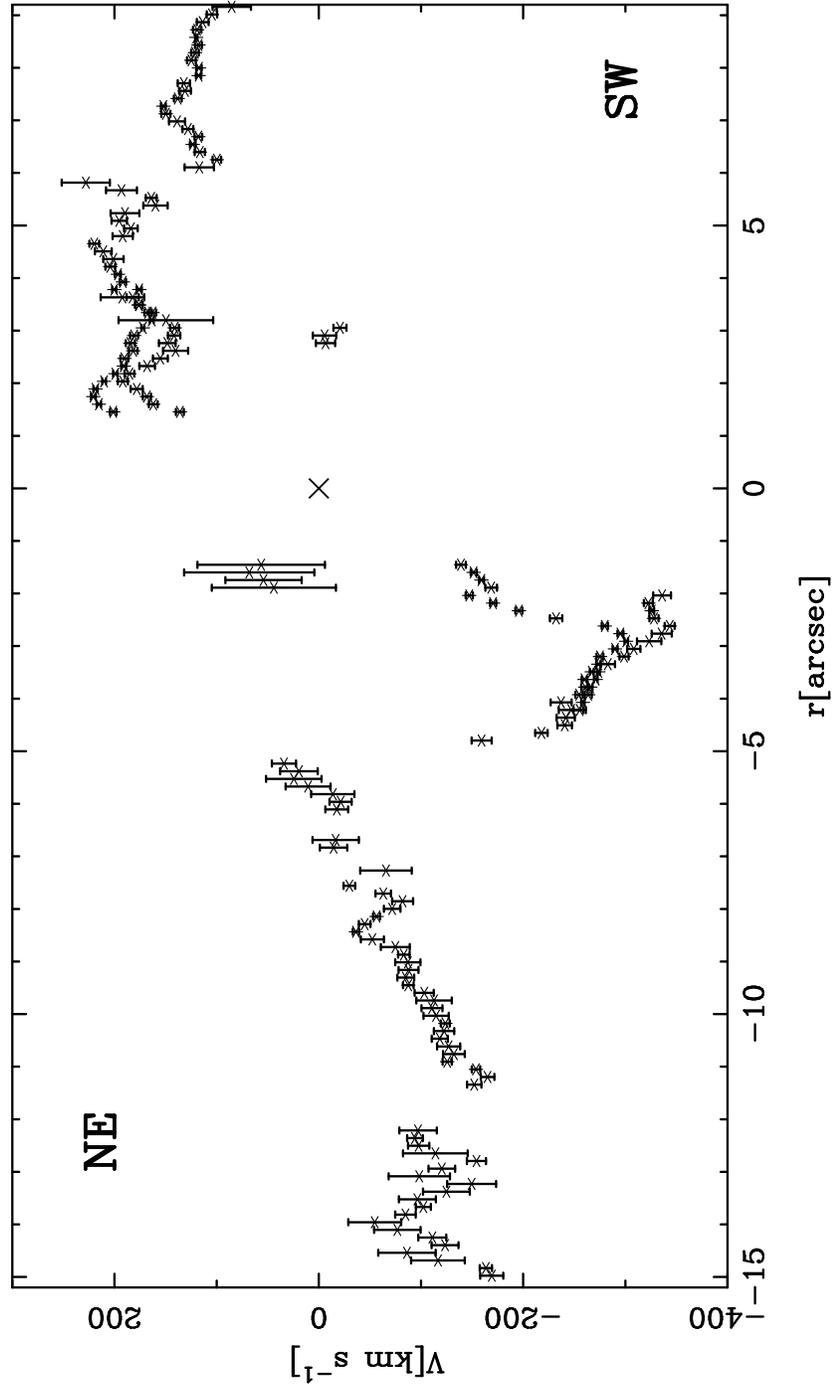}
\caption{Velocity field computed from the components that form the [O {\sc iii}]$\lambda5007$ emission-line profile. The cross indicates the galactic centre.\label{figure-2}}
\end{center}
\end{figure}

The complexity in the kinematics of the region has been already pointed out not only in the ionized gas but also in H{\sc i} \citep{mor05a}. So, as it was expected, the line profiles are kinematically complex in our Gemini spectrum too. We have extracted the 1-D spectra in a pixel-by-pixel way to avoid the loss of spatial information. The Gaussian decomposition is fully discused in \citet{rey13}. Up to three Gaussian components were needed in order to fit the [O{\sc iii}]$\lambda5007$ profile in the NE region. The structure of the SW region appears simpler than that of the NE in the long-slit spectrum (Fig.~\ref{figure-1}); actually, the NE region concentrates not only the strongest emission but also the most disturbed motion. However the Gaussian decomposition in the SW EELR revealed more than one component too. Fig.~\ref{figure-2} shows our velocity field which was computed from [O{\sc iii}]$\lambda5007$ decomposition. The velocities are expressed respect to the host galaxy systemic velocity. As the central region was excluded (Section~\ref{section:Data}) there are no data inside the inner 3~arcsec; the cross indicates the galactic centre. The most kinematical complexity, and therefore the most disturbed motion, appears towards the NE in agreement with the H$\alpha$ velocity field from \citet{mor05a}. Despite the 3$^{\circ}$-difference in the position angles with which our and their spectra were taken, we note that the high-velocity [O {\sc iii}] components are located between 2 and 5 arcsec, slightly further away from the nucleous than the H$\alpha$ high-velocity components are. While the rotation curve becomes stable at arount $\sim\pm$200~km~s$^{-1}$, the blue-shifted line-splitting in the NE region accounts for strongly disturbed movements that reach radial velocities of $\sim$400~km~s$^{-1}$.

The multiwavelength studies of 3C~305, the overlapping of its emission (or absortion, in the case of H{\sc i}), and the overall kinematics have pointed out that the radio jet is interacting with the ambient medium in a violent way. In the optical band, the main clue about such interaction have come from the kinematical point of view. The following section aims to show that the kinetical energy transfered by the jets into the ISM also drives shock waves that ionize the gas.

%
%

\section{The ionizing mechanism}
\label{section:discussion}
We are going to investigate whether the jet-cloud interaction can leave its trace in the ionization state of the region. The first spectral signature of shocks came from the [Fe {\sc ii}]$\lambda$1.644$\mu$m infrared (IR) emission line reported by \citet{jack03}. This line, first discovered in supernova remnants, is produced by collisional excitation with electrons. The strong shock waves triggered by the supernova explotion give raise to syncrotron emission within the recombination region. The iron line is excited by this same electron population, so near-IR and radio emission are overlapped in this cases \citep{moor88,kawa88,green91,forbes93}. The spatial relation between radio and [Fe {\sc ii}] emission in 3C~305 was shown by \citet{jack03}, who argued that the presence of shocks is evident at the jet termination point, where [Fe {\sc ii}] emission reaches its maximum. In addition, they have also shown that [Fe {\sc ii}] and [O {\sc iii}]$\lambda5007$ emission are overlapped and extended along the radio axis, although the oxygen emission is more extended. Since our Gemini spectrum was specially obtained in a direction roughly coincident with that of the radio axis, we are going to investigate whether the jets were able to trigger no only the [Fe {\sc ii}] emission but also the entire EELR emission through autoionizing shocks.

The pass of the radio jets drives shock waves within the medium where it propagates. As a result, the gas is heated and accelerated. The heating give raise to emission, as a cooling mechanism. And, since the clouds are accelerated, additional (secondary) shock waves might be created by the strong cloud-cloud interaction that occur within the clumpy medium \citep{heck82}. Although strongly dependent on velocities, the jets can also trigger auto-ionizing shocks. In such a case, those shocks will collisionally ionize the gas as well as photoionize the gas ahead of the shock. The contribution of this pre-ionized gas to the whole spectrum will depend on how fast the shock moves \citep{dop95,dop96,all08}. We want to analyse now the influence of a local radiation field created {\it in situ} by this kind of shock waves driven by the radio jet.

\begin{figure}
\begin{center}
\includegraphics[angle=270,scale=0.7]{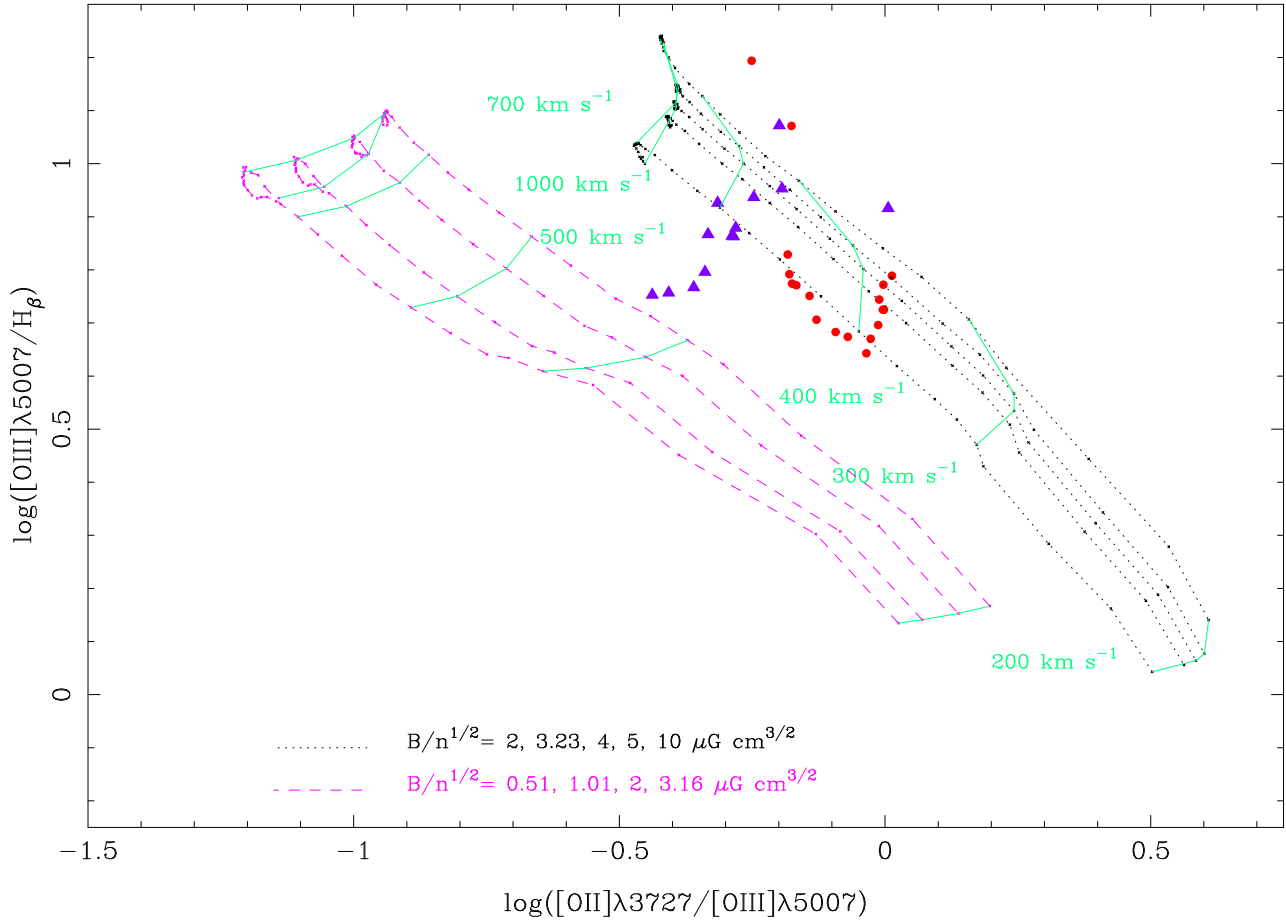}
\caption{Grids of shock-ionization (solar abundance) model \citep{all08} for two different ambient density: 100~cm$^{-3}$ as dotted lines, and 1000~cm$^{-3}$ as dashed lines. One line per magnetic parameter, the same values were used in both plots, note they differ depending upon density; the magnetic field strength was restricted to 16$\mu G<B<$100$\mu G$ (see text). The NE EELR is plotted as dots (red dots in the electronic version), and the SW EELR is plotted as triangles (violet triangles in the electronic version).}\label{figure-3}
\end{center}
\end{figure}

\begin{figure}
\begin{center}
\includegraphics[angle=0,scale=.8]{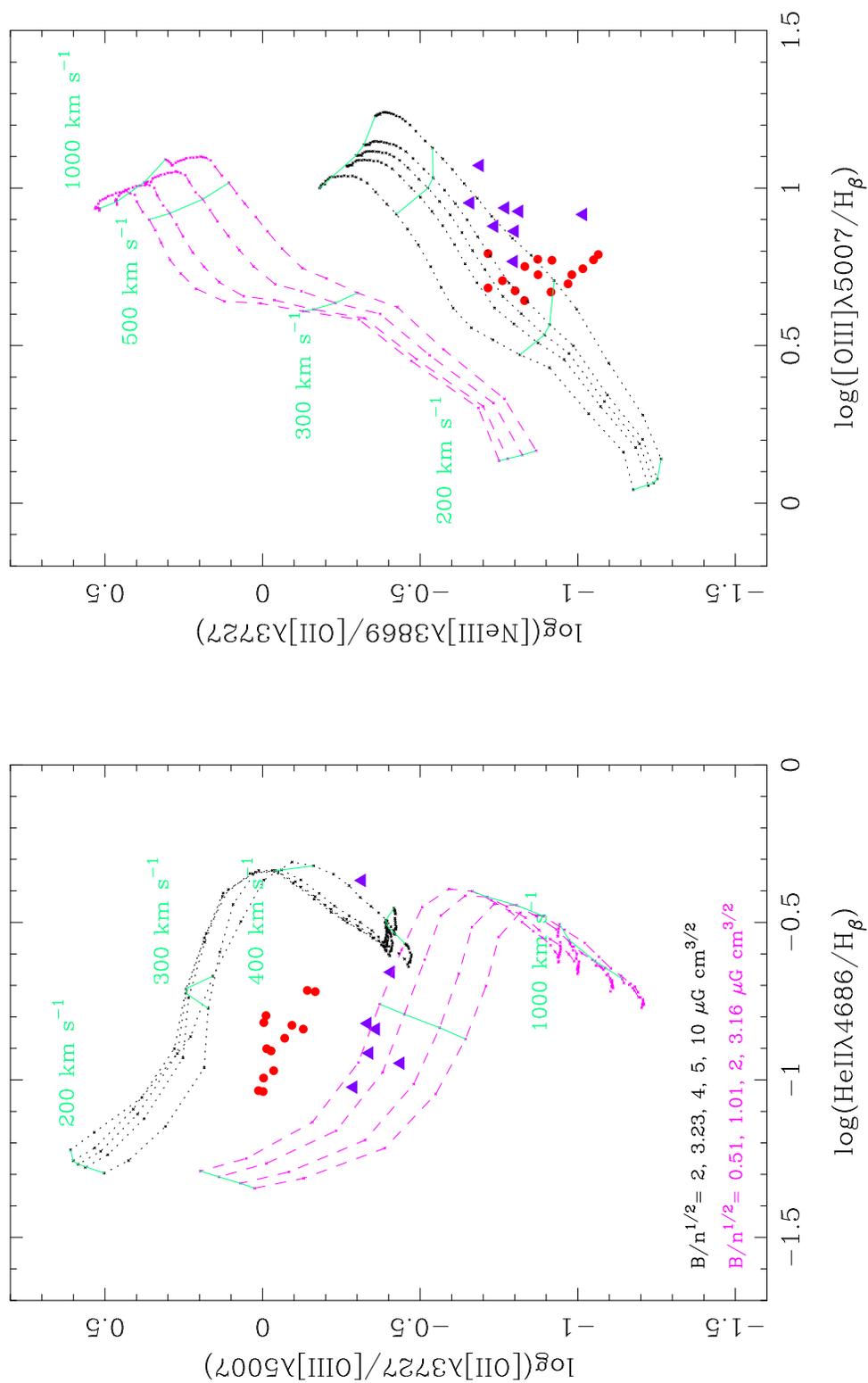}
\caption{Grids of solar abundance shock-ionization models. The [O{\sc ii}]$\lambda3727$/[O{\sc iii}]$\lambda5007$ as well as [Ne{\sc iii}]$\lambda3869$/H$\beta$ ratios are strongly sensitive to the pre-shock density when combined with [O{\sc iii}]$\lambda5007$/H$\beta$ and He{\sc ii}/H$\beta$. See also Fig.~\ref{figure-3} for model and data references.}\label{figure-4}
\end{center}
\end{figure}

\begin{figure}
\begin{center}
\includegraphics[angle=270,scale=0.9]{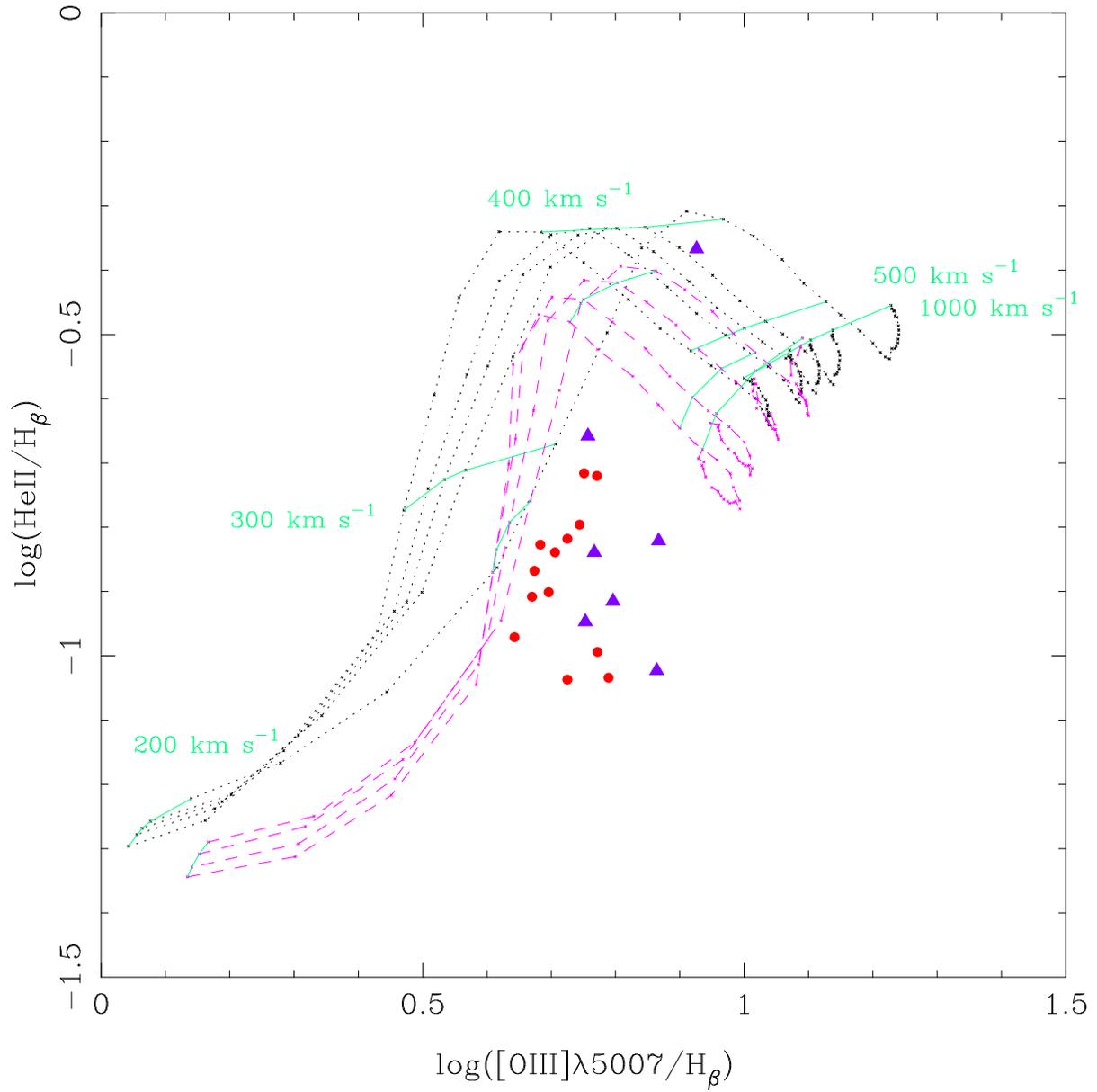}
\caption{Density information is lost when [O{\sc iii}]$\lambda5007$/H$\beta$ and He{\sc ii}/H$\beta$ ratios are combined in the same diagram. Models and data as in Fig.~\ref{figure-3}.}\label{figure-5}
\end{center}
\end{figure}

The shock-ionization models (solar abundances) from \citet{all08} were used in the diagnostic diagrams plotted in Figs.~\ref{figure-3}~to~\ref{figure-5}. We have plotted the shock+precursor grids (i.e. auto-ionizing shocks) for two values of pre-shock density $n$: 10$^2$~cm$^{-3}$ (dotted lines) and 10$^3$~cm$^{-3}$ (dashed lines). The ambient density has been constrained by using optical lines, but it has been calculated under photoionisaion conditions \citep[case B,][]{ost89}. \citet{mor05a} derived $n<$~500~cm$^{-3}$ by using [S {\sc ii}]$\lambda \lambda$6717,6731, but \citet{heck82} had previously found 1000~cm$^{-3}$ with the same line-ratio, altough some uncertainties may have affected the result. Since this is the only information we have regarding the optical band, we use the shock-ionization grids with 10$^2$ and 10$^3$~cm$^{-3}$ as extreme density values, although it is worthwhile noting that the density might be very different under shock-ionization conditions. The shock-only (non auto-ionizing) grids lie outside the region covered by our observations, so they were not plotted. Each curve is characterized by one {\it magnetic parameter}: $B/n^{1/2}$ value, where B is the magnetic field strength. \citet{har12} have constrained the magnetic field strength from radio depolarization and X-ray measurements, giving 16$<B<$100$\mu G$. We have traced the curves whose magnetic field lies in this range. The shock velocities are also pointed out (straight lines).

Each EELR was plotted separately, the NE region as dots, and the SW one as triangles. The diagrams combine high- and low-excitation lines, being He{\sc ii} and [Ne{\sc iii}]$\lambda3869$ the highest-excitation lines in the spectrum. The observations are well fitted by the models, regardeless of the way in which line-ratios are formed. The NE and SW data sets are clearly detached from each other, mostly because of their [O{\sc ii}]$\lambda3727$/[O{\sc iii}]$\lambda5007$ intensity.

The models reproduce the observed relative intensities if the shock propagates at 300$<v<$500~km~s$^{-1}$. The same disturbed velocities that we find in the velocity field. Lines such as [O{\sc ii}]$\lambda3727$ and H$\beta$ are particularly intense. These low ionization lines are usually formed within the recombination region behind the shock front. This is the place where the bulk of collisional excitation lines are formed. However, if we just consider the emission coming from the recombination region we should find significantly lower intensities, almost an order of magnitude lower than the observed in [O{\sc ii}]$\lambda3727$/[O{\sc iii}]$\lambda5007$ and [O{\sc iii}]$\lambda5007$/H$\beta$ ratios. It is also worth highlighting that, although different mean values characterise each EELR, the [O{\sc ii}]$\lambda3727$/[O{\sc iii}]$\lambda5007$ line-ratio keeps almost constant within each region (Fig.~\ref{figure-3} and Fig~\ref{figure-4}, left hand panel.). This is an important result since it implies that the ionizing conditions are not changing within the inner $\pm$4.5~arcsec ($\pm$4~kpc) approximately.

Regarding the [O{\sc ii}]$\lambda3727$/[O{\sc iii}]$\lambda5007$ and [Ne{\sc iii}]$\lambda3869$/[O{\sc ii}]$\lambda3727$ line-ratios (Figs.~\ref{figure-3}-\ref{figure-4}), both the theoretical and observed behaviour deserve special discusion. This will be provided in the following subsections of the paper.

\subsection{Why the EELR is not photoionized?}

We may wonder if we can explain such an enhancement in the [O{\sc ii}]$\lambda3727$ emission as an improvement on the nuclear ionizing power once the shocks have passed throughout the EELR clouds. Nuclear photoionization may certainly play a role, but it has to be of secondary relevance in the case of 3C~305, since it is hard to reconcile with these three important clues:\\
- the ionization state of the region does not drop with radial distance. It was stated by \citet{heck82} and is confirmed with our new data. The intensity profiles peak at 2-3 arcsec ($\sim$2~kpc) from the nucleus, including the high-excitation He{\sc ii}, [Ne{\sc iii}]$\lambda3869$, or [O{\sc iii}]$\lambda5007$ lines. But the most important clue in this regard is the way in which [O{\sc ii}]$\lambda3727$/[O{\sc iii}]$\lambda5007$ behaves. As we said before, it keeps quite constant, showing no evidence of spatial variation. If the ISM was photoionized, the geometrical dilution of photons should be mirrored by this line-ratio \citep[][and references therein]{pen90}; it means that spectral variation as a function of distance from the nucleus should be noticeable. In the same sense, the X-ray hardness ratio \citep{har12} neither shows spatial variation, which implies that no spectral variation can be inferred from this EM band either. In other words, the lack of ionizing photons with increasing distance, characteristic of photoionization, is not observed in the optical nor X-rays;\\
- the kinematics, which most disturbed components are found at almost the same location that the maximum line intensity;\\
- the X-ray extended emission was successfuly reproduced by collisional ionization. It is worth emphasizing at this point that both optical and soft X-ray emission may have the same origin \citep{bia06,mass09,har12}.

\subsection{The ambient density.}

The theoretical grids clearly show that the oxygen ratio as well as [Ne{\sc iii}]$\lambda3869$/[O{\sc ii}]$\lambda3727$, when combined with [O{\sc iii}]$\lambda5007$/H$\beta$ and He{\sc ii}/H$\beta$ are strongly sensitive to the pre-shock density (Figs.~\ref{figure-3}-\ref{figure-4}). None of these latter ratios by themselves provide any information about density when combined with each other (Fig.~\ref{figure-5}), but they are useful in separating the grids. The observations are best reproduced by low density models as a whole. 

An interesting result is found when [O{\sc ii}]$\lambda3727$/[O{\sc iii}]$\lambda5007$ and [Ne{\sc iii}]$\lambda3869$/[O{\sc ii}]$\lambda3727$ are combined in the same diagram. The grids overlap each other, so the diagram is completely degenerated in density \footnote{We do not show this diagram since it would be illegible. However, it can be easily built with the ITERA code from \citet{gro10}.}. Such a degeneracy can be broken up, for instance, with He{\sc ii}/H$\beta$ or [O{\sc iii}]$\lambda5007$/H$\beta$ in a three-dimensional diagram. Since the huge {\sc mapping~iii} data base \citep{all08} is comprised of several densities and abundaces data sets, we were able to test different kind of scenarios. Other combinations between density and metallicity, besides those used in the previous diagrams, were studied. These changes, and the reason why they were tested for, are explained in the next paragraph. The key point here is that if He{\sc ii}/H$\beta$ is used instead of [O{\sc iii}]$\lambda5007$/H$\beta$ the degeneracy is simultaneusly solved for density and abundance too, if shock velocities are slower than 700~km~s$^{-1}$.

\citet{har12} suggested that the X-ray plasma and the line-emitting gas may be in rough pressure balance. Under this hypothesis, the temperature in the ionized gas (T~$\gtrsim$~10$^4$~K) limits the density to the range $n\lesssim$~180~cm$^{-3}$ (this is a crude estimation since we did not perform any consideration regarding the X-ray nor line-emitting volumes). Nevertheless, it has to be highlighted that the X-ray temperature and density (from which these optical temperature and density relation is derived) were obtained for a sub-solar abundance gas. For this reason, we have examined the auto-ionizing shock-model data base in order to adapt the models to the physical conditions derived from pressure-balance hypothesis. 

\begin{figure}
\begin{center}
\includegraphics[scale=1]{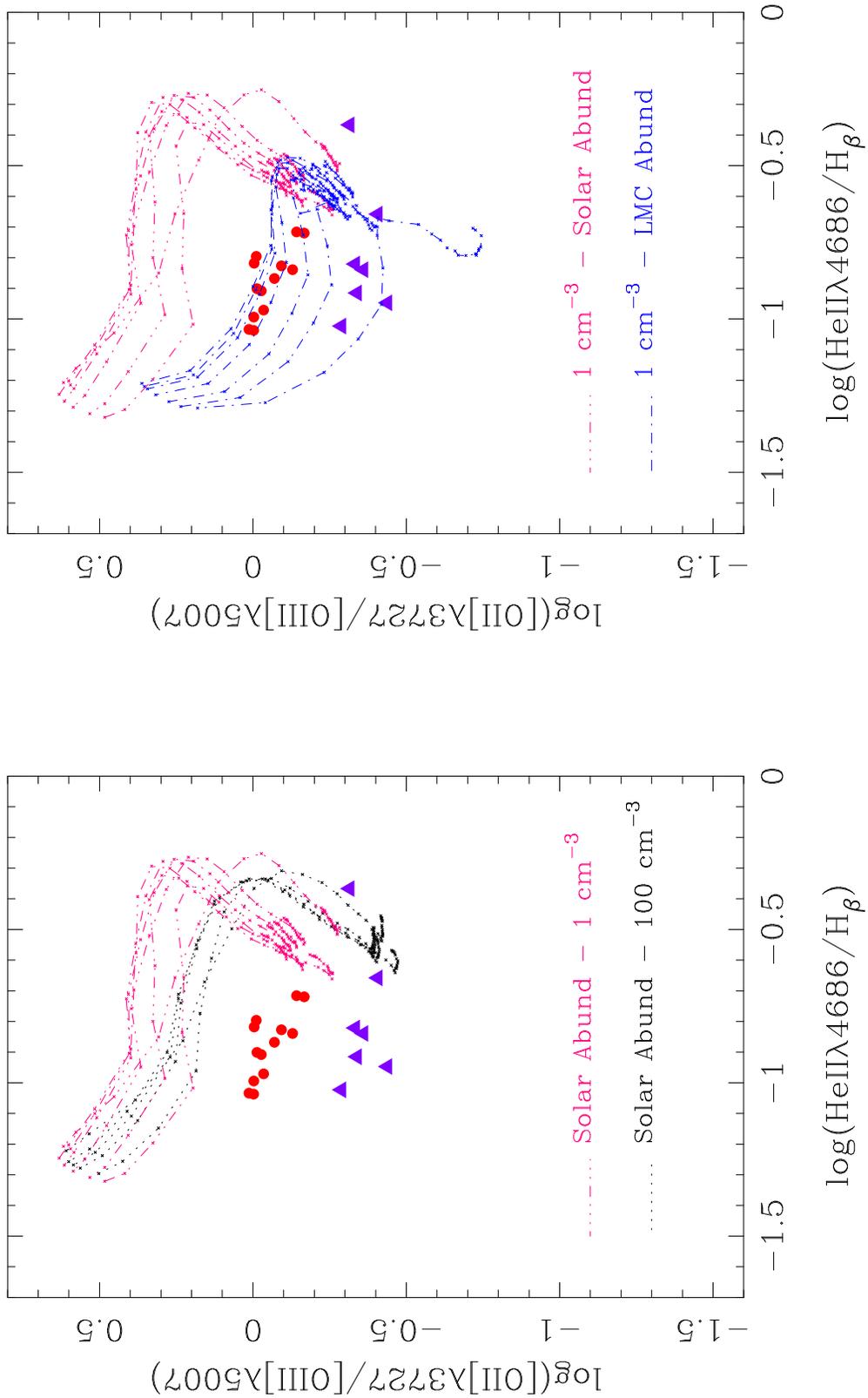}
\caption{Comparison of densities and abundances in shock-ionization models. Dotted lines: solar abundance, 100~cm$^{-3}$ (the same as previous diagrams); dot-dot-dashed lines: solar abundance, 1~cm$^{-3}$; dot-dashed lines: LMC abundance \citep[see][]{all08}, 1~cm$^{-3}$. {\it Left hand panel}: the abundance is fixed to the solar value, while the density varies. {\it Right hand panel}: the density is fixed to 1~cm$^{-3}$, while the abundance lowers. As in previous figures, the NE EELR is plotted as dots (red dots in the electronic version), and the SW EELR is plotted as triangles (violet triangles in the electronic version). Data are best reproduced by low-density and low-abundance models, which support the hypothesis that X-ray plasma is in pressure balance with the optical plasma (see text).}\label{figure-6}
\end{center}
\end{figure}

The projection onto [O{\sc ii}]$\lambda3727$/[O{\sc iii}]$\lambda5007$-He{\sc ii}/H$\beta$ plane (Fig.~\ref{figure-6}) is the most clear way of viewing that 3-D diagram we mentioned before ([O{\sc ii}]$\lambda3727$/[O{\sc iii}]$\lambda5007$ vs. [Ne{\sc iii}]$\lambda3869$/[O{\sc ii}]$\lambda3727$ vs. He{\sc ii}/H$\beta$). For simplicity, we have only drawn the way in which the density behaves when the abundance is fixed to the solar value (left-hand panel), and also the way in which the abundance behaves when the density is fixed to $n=$1~cm$^{-3}$ (right-hand panel). The density range tend to favour low densities, so we have chosen $n=$1~cm$^{-3}$  because it is the lowest density value in the library that let us perform the comparison. The following parameter combinations were tested:\\
- solar abundance - $n=$100~cm$^{-3}$ (dotted lines, as in previous diagrams)\\
- solar abundance - $n=$1~cm$^{-3}$ (dot-dot-dashed lines)\\
- Large Magellanic Cloud \citep[see][(LMC)]{all08} abundance - $n=$1~cm$^{-3}$ (dot-dashed lines)\\

The LMC abundance plays the role of the sub-solar abundance found by \citet{har12}. We did not plot the shock velocities, nor the magnetic parameters in order of simplifying the diagram. The same ranges as in previous diagrams were used. The left hand panel of Fig.~\ref{figure-6} shows that lowering the density is not enough in order to reproduce the line ratios if we keep the solar abundance. The observation are successfully reproduced when both the density and the abundance are lowered (Fig.~\ref{figure-6}, right hand panel), strongly supporting the pressure balance hypothesis. In addition, an ambient density of around 1~cm$^{-3}$ implies that the temperature should be T~$\sim$10$^6$~K in the ionized gas. Indeed, this is the characteristic temperature of the {\it high temperature radiative zone} and the {\it non equilibrium cooling zone} of the shock \citep[see][for a detailed description]{dop96,all08}, where the models predict the UV and soft X-ray emission.

\section{Summary and Conclusions.}
\label{section:summary}
A detailed analysis of the physical conditions of the EELR in 3C~305 was made. A large set of previous and self-consistent multiwavelenght studies have shown that the gas has been disturbed and also shock-heated by the interaction with the radio jet. We have analyse the ionization condition of the EELR's material following two steps. First, we discussed the shock-ionization models under density constraints obtained within the line-emittig gas. We found that the observations are better reproduced by low-density grids in the range 300$<v<$500~km~s$^{-1}$. The low-ionization lines in the spectrum are especially intense, which implies that the recombination region behind the shock front contributes significantly to the whole emission. However, in order to explaine the observed intensities, the contribution from the precursor region is crucial. Finally, based upon the X-ray parameters determined by \citet{har12}, and following the hypothesis that the X-ray plasma is in pressure balance with the EELR's gas, we explored how the models respond to changes in density and abundance. We found that the best fit to the observations is reached when shock waves propagate throughout a low-density and low-metallicity medium. The metallicity in our models (LMC) is comparable to that derived from the X-ray study (0.1-0.2~solar). The pressure balance hypothesis establish a relationship between density and temperature in the optical band, where our best-fit density (1~cm$^{-3}$) implies a rather high temperature ($\sim$10$^6$~K). However, such a temperature is characteristic of the recombination region just behind the shock front, where the bulk of low-ionization lines are formed. 

The enhancement in line-emission, shown by our new Gemini data, is directly explained by the interaction between the powerful jets and the ambient medium through strong shocks. The high amount of mechanical energy injected in the system drives very fast shock waves. The fastest shocks create a local radiation field whose power is directly related to the shock velocity: the faster the shock, the harder the ionizing spectrum  \citep[see fig.~1 from][]{all08}. The photons created {\it in situ} ionize the ambient medium before it is reached by the shock (\textquotedblleft the precursor''). The precursor is capable of producing a large amount of emission, in addition to that produced within the recombination region. In this sense, the effectiveness of cooling has to be highlighted by the presence of neutral outflow \citep{mor05a}.

The traces of the radio jets throughout the ISM are being unveiled now. The input of mechanical energy not only accelerates the EELR's clouds, but also plays a very important role in triggering the optical line-emission and presumably the soft X-ray emission too. This analysis intends to achieve the final step in understanding how the radio components relate to the optical emitting gas; we have found the optical missing piece in the whole picture that \citet{har12} described. The shock waves driven by the interaction between the radio jet and the ambient gas have had consequences in every EM band.

\bsp
%
%


%
%

\bibliographystyle{mn2e}
\bibliography{biblios.bib}

\begin{thebibliography}{}

\bibitem[\protect\citeauthoryear{{Allen}, {Groves}, {Dopita}, {Sutherland} \&
  {Kewley}}{{Allen} et~al.}{2008}]{all08}
{Allen} M.~G.,  {Groves} B.~A.,  {Dopita} M.~A.,  {Sutherland} R.~S.,
  {Kewley} L.~J.,  2008, \apjs, 178, 20

\bibitem[\protect\citeauthoryear{{Bianchi}, {Guainazzi} \&
  {Chiaberge}}{{Bianchi} et~al.}{2006}]{bia06}
{Bianchi} S.,  {Guainazzi} M.,    {Chiaberge} M.,  2006, \aap, 448, 499

\bibitem[\protect\citeauthoryear{{Christensen}, {Jahnke}, {Wisotzki},
  {S{\'a}nchez}, {Exter} \& {Roth}}{{Christensen} et~al.}{2006}]{chr06}
{Christensen} L.,  {Jahnke} K.,  {Wisotzki} L.,  {S{\'a}nchez} S.~F.,  {Exter}
  K.,    {Roth} M.~M.,  2006, \aap, 452, 869

\bibitem[\protect\citeauthoryear{{Dopita} \& {Sutherland}}{{Dopita} \&
  {Sutherland}}{1995}]{dop95}
{Dopita} M.~A.,  {Sutherland} R.~S.,  1995, \apj, 455, 468

\bibitem[\protect\citeauthoryear{{Dopita} \& {Sutherland}}{{Dopita} \&
  {Sutherland}}{1996}]{dop96}
{Dopita} M.~A.,  {Sutherland} R.~S.,  1996, \apjs, 102, 161

\bibitem[\protect\citeauthoryear{{Feinstein}, {Macchetto}, {Martel} \&
  {Sparks}}{{Feinstein} et~al.}{2002}]{fei02}
{Feinstein} C.,  {Macchetto} F.~D.,  {Martel} A.~R.,    {Sparks} W.~B.,  2002,
  \apj, 565, 125

\bibitem[\protect\citeauthoryear{{Feinstein}, {Macchetto}, {Martel}, {Sparks}
  \& {McCarthy}}{{Feinstein} et~al.}{1999}]{fei99}
{Feinstein} C.,  {Macchetto} F.~D.,  {Martel} A.~R.,  {Sparks} W.~B.,
  {McCarthy} P.~J.,  1999, \apj, 526, 623

\bibitem[\protect\citeauthoryear{{Forbes} \& {Ward}}{{Forbes} \&
  {Ward}}{1993}]{forbes93}
{Forbes} D.~A.,  {Ward} M.~J.,  1993, \apj, 416, 150

\bibitem[\protect\citeauthoryear{{Greenhouse}, {Woodward}, {Thronson} Jr.,
  {Rudy}, {Rossano}, {Erwin} \& {Puetter}}{{Greenhouse} et~al.}{1991}]{green91}
{Greenhouse} M.~A.,  {Woodward} C.~E.,  {Thronson} Jr. H.~A.,  {Rudy} R.~J.,
  {Rossano} G.~S.,  {Erwin} P.,    {Puetter} R.~C.,  1991, \apj, 383, 164

\bibitem[\protect\citeauthoryear{{Groves} \& {Allen}}{{Groves} \&
  {Allen}}{2010}]{gro10}
{Groves} B.~A.,  {Allen} M.~G.,  2010, \na, 15, 614

\bibitem[\protect\citeauthoryear{{Hardcastle}, {Massaro} \&
  {Harris}}{{Hardcastle} et~al.}{2010}]{har10}
{Hardcastle} M.~J.,  {Massaro} F.,    {Harris} D.~E.,  2010, \mnras, 401, 2697

\bibitem[\protect\citeauthoryear{{Hardcastle}, {Massaro}, {Harris}, {Baum},
  {Bianchi}, {Chiaberge}, {Morganti}, {O'Dea} \& {Siemiginowska}}{{Hardcastle}
  et~al.}{2012}]{har12}
{Hardcastle} M.~J.,  {Massaro} F.,  {Harris} D.~E.,  {Baum} S.~A.,  {Bianchi}
  S.,  {Chiaberge} M.,  {Morganti} R.,  {O'Dea} C.~P.,    {Siemiginowska} A.,
  2012, \mnras, 424, 1774

\bibitem[\protect\citeauthoryear{{Heckman}, {Miley}, {Balick}, {van Breugel} \&
  {Butcher}}{{Heckman} et~al.}{1982}]{heck82}
{Heckman} T.~M.,  {Miley} G.~K.,  {Balick} B.,  {van Breugel} W.~J.~M.,
  {Butcher} H.~R.,  1982, \apj, 262, 529

\bibitem[\protect\citeauthoryear{{Jackson}, {Beswick}, {Pedlar}, {Cole},
  {Sparks}, {Leahy}, {Axon} \& {Holloway}}{{Jackson} et~al.}{2003}]{jack03}
{Jackson} N.,  {Beswick} R.,  {Pedlar} A.,  {Cole} G.~H.,  {Sparks} W.~B.,
  {Leahy} J.~P.,  {Axon} D.~J.,    {Holloway} A.~J.,  2003, \mnras, 338, 643

\bibitem[\protect\citeauthoryear{{Kawara}, {Taniguchi} \& {Nishida}}{{Kawara}
  et~al.}{1988}]{kawa88}
{Kawara} K.,  {Taniguchi} Y.,    {Nishida} M.,  1988, \apjl, 328, L41

\bibitem[\protect\citeauthoryear{{M{\'a}rquez}, {P{\'e}contal}, {Durret} \&
  {Petitjean}}{{M{\'a}rquez} et~al.}{2000}]{mar00}
{M{\'a}rquez} I.,  {P{\'e}contal} E.,  {Durret} F.,    {Petitjean} P.,  2000,
  \aap, 361, 5

\bibitem[\protect\citeauthoryear{{Massaro}, {Chiaberge}, {Grandi},
  {Giovannini}, {O'Dea}, {Macchetto}, {Baum}, {Gilli}, {Capetti}, {Bonafede} \&
  {Liuzzo}}{{Massaro} et~al.}{2009}]{mass09}
{Massaro} F.,  {Chiaberge} M.,  {Grandi} P.,  {Giovannini} G.,  {O'Dea} C.~P.,
  {Macchetto} F.~D.,  {Baum} S.~A.,  {Gilli} R.,  {Capetti} A.,  {Bonafede} A.,
     {Liuzzo} E.,  2009, \apjl, 692, L123

\bibitem[\protect\citeauthoryear{{Moorwood} \& {Oliva}}{{Moorwood} \&
  {Oliva}}{1988}]{moor88}
{Moorwood} A.~F.~M.,  {Oliva} E.,  1988, \aap, 203, 278

\bibitem[\protect\citeauthoryear{{Morganti}, {Oosterloo}, {Tadhunter}, {van
  Moorsel} \& {Emonts}}{{Morganti} et~al.}{2005}]{mor05a}
{Morganti} R.,  {Oosterloo} T.~A.,  {Tadhunter} C.~N.,  {van Moorsel} G.,
  {Emonts} B.,  2005, \aap, 439, 521

\bibitem[\protect\citeauthoryear{{Osterbrock}}{{Osterbrock}}{1989}]{ost89}
{Osterbrock} D.~E.,  1989, {Astrophysics of gaseous nebulae and active galactic
  nuclei}

\bibitem[\protect\citeauthoryear{{Penston}, {Robinson}, {Alloin},
  {Appenzeller}, {Aretxaga}, {Axon}, {Baribaud}, {Barthel}, {Baum}, {Boisson}
  \& et al.}{{Penston} et~al.}{1990}]{pen90}
{Penston} M.~V.,  {Robinson} A.,  {Alloin} D.,  {Appenzeller} I.,  {Aretxaga}
  I.,  {Axon} D.~J.,  {Baribaud} T.,  {Barthel} P.,  {Baum} S.~A.,  {Boisson}
  C.,    et al. 1990, \aap, 236, 53

\bibitem[\protect\citeauthoryear{{Reynaldi} \& {Feinstein}}{{Reynaldi} \&
  {Feinstein}}{2013}]{rey13}
{Reynaldi} V.,  {Feinstein} C.,  2013, \mnras, 430, 2221

\bibitem[\protect\citeauthoryear{{Sol{\'o}rzano-I{\~n}arrea}, {Tadhunter} \&
  {Axon}}{{Sol{\'o}rzano-I{\~n}arrea} et~al.}{2001}]{sol01}
{Sol{\'o}rzano-I{\~n}arrea} C.,  {Tadhunter} C.~N.,    {Axon} D.~J.,  2001,
  \mnras, 323, 965

\bibitem[\protect\citeauthoryear{{Sol{\'o}rzano-I{\~n}arrea}, {Tadhunter} \&
  {Bland-Hawthorn}}{{Sol{\'o}rzano-I{\~n}arrea} et~al.}{2002}]{sol02}
{Sol{\'o}rzano-I{\~n}arrea} C.,  {Tadhunter} C.~N.,    {Bland-Hawthorn} J.,
  2002, \mnras, 331, 673

\bibitem[\protect\citeauthoryear{{Tilak}, {O'Dea}, {Tadhunter}, {Wills},
  {Morganti}, {Baum}, {Koekemoer} \& {Dallacasa}}{{Tilak} et~al.}{2005}]{til05}
{Tilak} A.,  {O'Dea} C.~P.,  {Tadhunter} C.,  {Wills} K.,  {Morganti} R.,
  {Baum} S.~A.,  {Koekemoer} A.~M.,    {Dallacasa} D.,  2005, \aj, 130, 2513

\bibitem[\protect\citeauthoryear{{van Dokkum}}{{van Dokkum}}{2001}]{dok01}
{van Dokkum} P.~G.,  2001, \pasp, 113, 1420

\end{thebibliography}
\newpage




\begin{table}
\begin{center}
\begin{minipage}{340mm}
\caption{Set of emission-lines in the EELR of 3C 305. Fluxes are relative to that of H$\beta$ for each selected position.}
\label{table-1}
\begin{tabular}{lcccccc}
\hline
Line (\AA) & \multicolumn{1}{c}{Flux 3.8'' NE} & \multicolumn{1}{c}{Flux 2.6'' NE} & \multicolumn{1}{c}{Flux 1.5'' NE} & \multicolumn{1}{c}{Flux 1.5'' SW} & \multicolumn{1}{c}{Flux 2.6'' SW} & \multicolumn{1}{c}{Flux 3.3'' SW} \\
\hline
\hline

[O {\sc ii}]$\lambda3727$ & 4.1 & 4.39 & 8.34 & 4.08 & 3.96 & 15.66 \\

[Ne {\sc iii}]$\lambda3869$ & 0.79 & 0.53 & 0.8 & 0.63 & 0.73 & - \\

He {\sc ii} (4686) & - & 0.12 & - & 0.43 & - & - \\

H$\beta$ (4861) & 1 & 1 & 1 & 1 & 1 & 1 \\

[O {\sc iii}]$\lambda4959$ & 2.08 & 1.52 & 2.91 & 2.74 & 2.53 & 6.78 \\

[O {\sc iii}]$\lambda5007$ & 6.2 & 4.67 & 8.24 & 8.44 & 7.57 & 20.1 \\

[N {\sc i}]$\lambda5200$ & 0.19 & 0.26 & 0.36 & 0.51 & 0.26 & - \\

\hline
\end{tabular}
\end{minipage}
\end{center}
\end{table}

\newpage
%


%

\label{lastpage}

\end{document}